\documentclass[12pt,onecolumn,twoside]{JCTA}

\usepackage{graphicx}% or use the graphicx package for more complicated function

\usepackage{multicol}% for the format of two columns or more

\usepackage{amsmath}
\usepackage{amsthm}
\usepackage{amsfonts}
\usepackage{amssymb}% provides various useful mathematical symbols
\usepackage{mathrsfs}

\usepackage{lettrine}% for format of authors' biography

\usepackage[sort]{cite}

\usepackage{times}% for defining font as times

\usepackage{booktabs}% for creating table

\usepackage{wasysym}

\usepackage{fancyhdr}

\usepackage{ulem}
\usepackage{multirow}
\allowdisplaybreaks

\newtheoremstyle{mystyle}{1pt}{1pt}{\normalfont}{\parindent}{\bfseries}{}{1em}{}
\theoremstyle{mystyle}

\renewcommand{\footnotesep}{1pt}
\renewcommand{\footnotesize}{\scriptsize}

\def\ubar#1{\underset{\raise0.3em\hbox{$\smash{\scriptscriptstyle-}$}}{#1}}

\makeatletter
\def\@compress@cite#1{%  % This is executed for each number
  \advance\@tempcnta\@ne % Now \@tempcnta is one more than the previous number
  \ifnum #1=\@tempcnta   % Number follows previous--hold on to it
        \def\@h@ld{\citedash \citeform{#1}}%
  \else   %  non-successor -- dump what's held and do this one
     \@h@ld \@citea \citeform{#1}%
     \let\@h@ld\@empty
  \fi \@tempcnta#1\let\@citea\citepunct
} \makeatother
\begin{document}                          % Please begin your paper here.
\makeatletter
\def\@autr{{J. Ren} et al.}             % Change the primary author name here. Has it capitalized.
\makeatother

\begin{frontmatter}                       % The frontmatter environment contains Title, authors and
                                          % addresses; Abstract; Keywords

                                          % Title and footnote, please set suitable length of every line
\title{Optimal Control Theory in Intelligent Transportation Systems Research - A Review}\footnotetext
%%%%%%%%%%%%%%%%%%%%%%%%%%%%%%

\author[1]{Jimmy SJ. REN}$^{\dag}$,{ }
\author[1]{Wei WANG},{ }
\author[1]{Stephen Shaoyi LIAO}

\address[1]{.Department of Information Systems,
City University of Hong Kong, 83 Tat Chee Ave, Kowloon, Hong Kong SAR,
China}


\begin{abstract}
Continuous motorization and urbanization around the globe leads to an expansion of population in major cities. Therefore, ever-growing pressure imposed on the existing mass transit systems calls for a better technology, Intelligent Transportation Systems (ITS), to solve many new and demanding management issues. Many studies in the extant ITS literature attempted to address these issues within which various research methodologies were adopted. However, there is very few paper summarized what does optimal control theory (OCT), one of the sharpest tools to tackle management issues in engineering, do in solving these issues. It{\textquoteright}s both important and interesting to answer the following two questions. (1) How does OCT contribute to ITS research objectives? (2) What are the research gaps and possible future research directions? We searched 11 top transportation and control journals and reviewed 41 research articles in ITS area in which OCT was used as the main research methodology. We categorized the articles by four different ways to address our research questions. We can conclude from the review that OCT is widely used to address various aspects of management issues in ITS within which a large portion of the studies aimed to reduce traffic congestion. We also critically discussed these studies and pointed out some possible future research directions towards which OCT can be used.
\end{abstract}

\end{frontmatter}

\section{Introduction}
Conventional transportation systems are experiencing significant and ever-growing pressure during the past several decades due to the continuous motorization and urbanization around the world. While continuous motorization and urbanization lead to the increase of urban population, demand for automobile is also growing rapidly. It is not uncommon that various traffic issues such as traffic congestion, traffic accident and environmental issues become much more serious in metropolises. In addition to the traditional methods of ``solving" these problems, namely building more infrastructures, an emerging technology called Intelligent Transportation Systems (ITS) provides an alternative which is creating far-reaching impact on solving current transportation problems. ITS equips the elements within the transportation systems, such as vehicles, roads, traffic lights, traffic signs, etc. with information technology (IT), so that they would become more intelligent. IT also empowers these elements to communicate with each other to make more systematic decisions to reduce the potential conflict or error which may cause transportation problems. In many countries where embrace ITS, it has brought significant improvements in various ways including reduced congestion, enhanced safety and traveler satisfaction and convenience [1].

Due to the nature of ITS, a large number of research topics in this area could be formulated as problems of finding optimal operation policies. In other words, the underlying problem is how to find optimal policies to manage the transportation systems. For example, [2] provides an optimal policy to control the freeway ramp so that the goal of reducing traffic congestion is achieved. [3] manipulated the traffic signal timing policy to get the optimal result of congestion reduction. Similar examples can be found in more ITS research papers [4, 5, 6]. Many research topics in ITS share the same characteristic, namely the transportation systems act as a control system in which the inputs and the expected outputs can be clearly defined. It is of great value to find the optimal policies of operating the inputs to approach the expected output so that the whole system can be effectively managed. Not surprisingly, such problem domain fits nicely to the territory of optimal control theory (OCT), one of the most widely used research methodologies in addressing management issues in engineering.

Optimal control theory deals with the problem of finding a control law for a given system such that a certain optimality criterion is achieved. It is a branch of mathematics based on calculus of variations developed to find optimal ways to control a dynamic system. It{\textquoteright}s widely used in engineering, information systems and management science to solve various types of management problems [7]. In addition to our previous discussion, many other evidences could be found in the literature to testify that optimal control is suitable to solve research problems in ITS area [8, 9, 10].

Since OCT fits the context of ITS research, we believe it is both important and interesting to gain insights of the current adoption of this methodology in the contemporary ITS research. We attempt to answer the following research questions in this paper. First of all, how does OCT contribute to ITS research objectives? In other words, is OCT widely used and what particular research problems does optimal control solve in ITS? Second, what are the current research gaps and possible future research directions in ITS in which OCT can be effectively used? To our knowledge, there is very little prior studies addressed these research questions in the extant literature though we believe answering these research questions is important.

The rest of this paper is structured as the follows. Section two introduces the literature review methodology and categorizes the articles by 4 different ways so that our research questions can be properly addressed. Section three summarizes the findings and identified the current research gaps and points out the possible future research directions based on our review. Section four ends the paper with the concluding remarks.

\section{Literature review}
\subsection{Review methodology}
Our target journals include 11 top academic journals in transportation research, ITS research and automatic control. In the review, we only selected the research articles used OCT as their main research methodology. The studies should also directly address ITS problems or are strongly related to ITS. We mainly focus on relatively recent studies; however, some older research articles which are closely related to ITS topics are also included in our list.

We searched in these 11 journals by the key words ``optimal control", ``optimal control theory" in abstracts, keywords and titles, we then manually eliminated the articles without our review scope, e.g. doesn{\textquoteright}t address ITS problems. We finally obtained 41 research articles which fulfill our requirement.

\subsection{Categorize articles by academic journals}
The following table (Table 1) lists the number of selected research articles in each journal respectively. We can discern that two journals, ``Transportation Research Part B: Methodological" and ``Transportation Research Part C: Emerging Technologies" excels other ones in terms of article numbers. Furthermore, we can also observe that half of the articles we found in ``Transportation Research Part C: Emerging Technologies" were published in the past 5 years among which many of them were published very recently.

Consider the nature of these two journals by quoting the scope of them from their official website, ``Transportation Research Part B publishes papers on all methodological aspects of the subject … with important aspects of the design and/or analysis of transportation systems.", ``The focus of Transportation Research: Part C is high-quality, scholarly research that addresses development, applications, and implications, in the field of transportation, of emerging technologies …". This leads to the following interpretation of this phenomenon. As a relatively new research area, ITS is considered an emerging field with high potential impact in academia in which OCT is one of the widely adopted methodologies. Furthermore, according to the scope of Transportation Research Part B, we tend to believe that OCT is regarded as one of the rigorous and effective methodologies to solve ITS issues. However, it{\textquoteright}s still not clear in this stage that if OCT is capable of effectively solving different types of ITS problems. The next several sections will reveal this question.

\begin{center}
{Table 1~~Categorize articles by journals.}\vskip 3pt
\small{\begin{tabular}{@{ }lc@{ }} \toprule
Journals & Article No. \\\midrule
Transportation Research Part A: \\Policy and Practice (General) & 4\\ \hline 
Transportation Research Part B: \\Methodological  & 10\\ \hline 
Transportation Research Part C: \\Emerging Technologies  & 11\\ \hline 
Transportation Research Part D: \\Transport and Environment & 1\\ \hline 
Transportation Research Part E: \\Logistics and Transportation Review & 3\\ \hline 
Transportation Science & 4\\ \hline 
IEEE Transactions on Intelligent Transportation \\Systems & 4\\ \hline 
IEEE Transactions on Vehicular Technology & 1\\ \hline 
IEEE Transactions on Automatic Control & 1\\ \hline 
IEEE Intelligent Systems & 1\\ \hline 
IEEE Circuits and Systems Magazine & 1\\ \hline 
Total & 41\\
\bottomrule
\end{tabular}}
\vskip 3pt
\end{center}

\subsection{Categorize articles by traffic types in the research}
ITS doesn{\textquoteright}t limit the type of the traffic, thus traffic on highways, regular roads and even air traffic are within the scope of ITS. During the further analysis of the articles, we found there are four major traffic types the studies address, namely traffic on highways (freeways, motorways), traffic on regular roads in urban area, traffic on rail ways and air traffic. The following table (table 2) categorizes the articles by these four traffic types.

Though there are studies address ITS problems based on railway traffic and air traffic, we can easily tell from the table that ITS issues on land traffic dominate the studies in which OCT is used. This finding is consistent with the current main objectives of ITS, namely ``to improve the performance of highway, transit, and even air and maritime transportation systems" [1]. We can also learn from the table, although it is not obvious, that most of the studies solve problems for roads in urban area also address, either explicitly or implicitly, the same problem for highways. However, highways seem to have their own particular problem domain that regular roads don{\textquoteright}t share.

Take a closer scrutiny of the literature we found both highways/freeways/motorways and regular roads in urban area share the following ITS research topics which are likely to be addressed by OCT. They are ITS infrastructure maintenance and construction [11, 12, 13, 14], dynamic pricing [15, 16, 17, 18], dynamic traffic assignment [4, 19, 20], intelligent vehicle and energy control [21, 22] and incident management [23].

On the other hand, traffic in urban area embraces some studies which papers for highways don{\textquoteright}t share, e.g. traffic light control [34, 35, 36] and bus schedule control [37]. In addition to that, research topics addressed by OCT for highways, freeways or motorways only are the following. They are highway fuel economy [5, 30], highway pavement maintenance [6, 32] and highway speed limits and ramp metering [2, 3, 24, 25, 26, 27, 28, 29, 31, 33]. It is clear that there{\textquoteright}s a significant body of studies addresses ramp metering and speed limit. The reason is such research topics have a strong connection to traffic congestion reduction which is one of the most important goals of ITS. We will explicitly point that out in the later sections.

\newcommand{\minitab}[2][l]{\begin{tabular}{#1}#2\end{tabular}} 
\begin{center}
{Table 2~~Categorize articles by traffic type in the research.}\vskip 3pt
\small{\begin{tabular}{lll@{ }llc@{ }} \toprule
Traffic type & Articles \\\midrule
\multirow{3}{*}{\minitab[l]{Traffic on highways, \\ freeways or motorways}}
& [2, 3, 4, 5, 6, 11, 12, 13, 14, 15,\\ 
& 16,17,18,19, 20, 21, 22, 23, 24,\\
& 25, 26, 27, 28, 29, 30, 31, 32, 33]\\
\hline 
\multirow{2}{*}{\minitab[l]{Traffic in urban area} }
& [4, 11, 12, 13, 14, 15, 16, 17, 18,\\ 
& 19, 20, 21, 22, 23, 34, 35, 36, 37]\\
\hline 
\multirow{1}{*}{\minitab[l]{Traffic on railways} }
& [38, 39, 40, 41, 42, 43]\\ 
\hline 
\multirow{1}{*}{\minitab[l]{Air traffic} }
& [44, 45, 46]\\ 
\bottomrule
\end{tabular}}
\vskip 3pt
\end{center}

It{\textquoteright}s worth mentioning that many of the studies for rail traffic contributed to a more energy efficient system [38, 39, 41, 42] and most of the studies for air traffic focused on safety issues [44, 45]. Though rail traffic and air traffic are not at the core of ITS and the number of studies in these areas are relatively limited, the capability of OCT in solving important problems in these traffic types should not be overlooked.

In sum, we can summarize from table 2 that OCT is widely used in academic studies to address ITS research topics for different traffic types, among which the issues on land road is the current focus.

\subsection{Categorize articles by ITS research objectives}
In the previous section, we not only categorized papers by traffic types, we also summarized different ITS research topics for each traffic type. It{\textquoteright}s worthwhile to point out that ITS research topics don{\textquoteright}t necessarily represent the research objectives of ITS. For instance, dynamic pricing is a typical ITS research topic (or called ITS methodology), however one of the most important research goals of doing dynamic pricing in ITS is to reduce traffic congestion. It is reducing congestion the ITS research objective, not dynamic pricing. It{\textquoteright}s important to distinguish these two concepts because when we say OCT contributes to ITS research, we need to make sure OCT contributes to ITS research objectives, not just to methodologies.

As we mentioned previously, we are especially interested in the question ``How does OCT contribute to ITS research objectives". In order to answer this question, we further examined the literature and categorized them by research objectives in the following table (table 3).

It{\textquoteright}s not difficult to observe that a large number of (15 out of 41) studies talked about mitigating traffic congestions. We may conclude from this finding that OCT contributes to ITS research objectives most by addressing the issue of traffic congestion. This finding is interesting because it indicates that OCT is the methodology which is widely used to solve one of the major problems in contemporary transportation, namely reducing traffic congestion. It further implies that the usage of OCT in the extant literature is consistent with one of the main management objectives of ITS [1]. The second largest body of research addresses efficiency improvement. We can see that this body of studies is growing rapidly in recent years since many of the papers were published very recently. Meanwhile, various maintenance issues in ITS were also addressed by OCT. In addition, it is also not uncommon for optimal control to solve traffic safety issues.\\\\\\

\begin{center}
{Table 3~~Categorize articles by ITS research objectives.}\vskip 3pt
\small{\begin{tabular}{lll@{ }llc@{ }} \toprule
Research objectives & Articles \\\midrule
\multirow{2}{*}{\minitab[l]{Reduce traffic congestion}}
& [2, 3, 16, 17, 18, 24, 25, 26,\\
& 27, 28, 29, 31, 33, 35, 36]\\
\hline 
\multirow{1}{*}{\minitab[l]{Improve environment} }
& [15, 34]\\ 
\hline 
\multirow{2}{*}{\minitab[l]{Improve efficiency,\\e.g. reduce cost, or \\reduce energy consumption} }
& [5, 22, 30, 37, 38, 39, 41, 42,\\
& 46] \\ \\
\hline 
\multirow{1}{*}{\minitab[l]{Improve maintenance\\or construction} }
& [6, 11, 12, 13, 14, 32]\\ \\
\hline 
\multirow{1}{*}{\minitab[l]{Improve traffic forecasting\\and prediction} }
& [4, 19, 20]\\ \\
\hline 
\multirow{1}{*}{\minitab[l]{Improve safety} }
& [21, 23, 40, 43, 44, 45]\\
\bottomrule
\end{tabular}}
\vskip 3pt
\end{center}

In traffic congestion problem domain, various methods are used to solve the issue from different angles. For instance, [16] developed two types of dynamic congestion pricing models based on OCT to achieve equilibrium for commuters under the tolls. [31] presents a stochastic optimal control based approach to real-time incident responsive coordinated ramp control, etc. The next section discusses what detailed ITS methodologies are used to reduce traffic congestion in the extant literature by using OCT.

\subsection{Categorize articles by research methods of reducing traffic congestion}
It{\textquoteright}s important to gain more insights on reducing traffic congestion by using OCT because congestion reduction is essential to ITS [1]. Moreover, there is a considerable body of optimal control based studies within the scope of this subject. This section reveals what underlying techniques are adopted in the literature to reduce traffic congestion in which optimal control theory was used as the fundamental enabler of such techniques. Table 4 summarizes the literatures.

We found there are five methods in ITS to mitigate traffic congestion problems, they are dynamic pricing, ramp metering, variable speed limits, traffic light control and the combination of different methods, namely the integrated control approach. We can observe from table 4 that research articles we reviewed distribute to these categories pretty well. In addition, it{\textquoteright}s not difficult to see that more researchers are interested in using optimal control to address ramp metering problems. This implies that more papers were devoted to congestion reduction on highways, freeways and motorways. This makes sense, because on one hand the phenomenon of traffic congestion on highways is indeed quite serious in many places, on the other hand highways, freeways and motorways are relatively easier to model than regular roads in urban areas. This largely makes the optimal control problem more tractable and more realistic [1, 24].

\begin{center}
{Table 4~~Categorize articles by congestion reduction methods.}\vskip 3pt
\small{\begin{tabular}{lll@{ }llc@{ }} \toprule
Methods & Articles \\\midrule
\multirow{2}{*}{\minitab[l]{Dynamic pricing}} 
& [2, 3, 24, 25, 26, 27, 28, 31,\\
& 33] \\
\hline 
\multirow{1}{*}{\minitab[l]{Variable speed limits} }
& [24, 25, 26, 27]\\ 
\hline 
\multirow{1}{*}{\minitab[l]{Traffic light control} }
& [35, 36]\\
\hline 
\multirow{1}{*}{\minitab[l]{Integrated control \\approach(combines \\two or more methods)} }
& [6, 11, 12, 13, 14, 32]\\ \\ \\
\bottomrule
\end{tabular}}
\vskip 3pt
\end{center}

It{\textquoteright}s also worth pointing out that though there are not as many studies as that in addressing congestion reduction on highways, studies in reducing traffic congestion in urban roads should not be neglected. There are two main optimal control based methods in these studies, namely dynamic pricing and traffic light control in this area.

An integrated approach prevails in such studies by considering two or even three of these methods simultaneously to get a better result. For instance, Carlson et al. [24] combines the factors of ramp metering and variable speed limits and proposed mainstream traffic flow control method to improve traffic flow efficiency. They published another research article at the same year aiming to reduce the congestion in the bottleneck of highways by using a similar approach [25]. Other studies using the integrating approach can also be found in the literatures [26, 27, 29].

\subsection{Categorize articles by areas in ITS research}
Intelligent transportation systems include a wide and growing suite of technologies and applications. ITS area itself can be roughly grouped within five broad categories: (1) Advanced Traveller Information Systems; (2) Advanced Transportation Management Systems; (3) ITS-Enabled Transportation Pricing Systems; (4) Advanced Public Transportation Systems; (5) Fully Integrated Intelligent Transportation Systems [9]. Advanced Traveler Information Systems provide drivers with real-time traffic information including optimal routes, navigations, traffic congestion predictions, accident warnings, etc. to support the drivers to make better driving decisions. Advanced Transportation Management Systems involve traffic control devices to resolve traffic issues, such as traffic signals, ramp meters, variable message signs, and traffic operations centres. ITS-Enabled Transportation Pricing Systems mainly focus on the optimal congestion pricing. Advanced Public Transportation Systems allow trains and buses to report their positioning information to the passengers so passengers can be informed of their real-time arrival or departure information. Finally, Fully Integrated Intelligent Transportation systems include vehicle-to-infrastructure (V2I) and vehicle-to-vehicle (V2V) integration which enable assets to communication with each other in the transportation systems. Particular cases could be communications from vehicles to roadside sensors, traffic lights, and other vehicles.

We believe the aforementioned five areas well represent the five development trends of ITS in the future. Therefore research opportunities could be yielded from new insights in these areas. We try to put the articles we reviewed into these categories so that current research trend, research gaps and possible future research trend of using OCT in ITS may be discovered (see table 5).

\begin{center}
{Table 5~~Categorize articles by areas in ITS research.}\vskip 3pt
\small{\begin{tabular}{lll@{ }llc@{ }} \toprule
Methods & Articles \\\midrule
\multirow{1}{*}{\minitab[l]{Advanced Traveler\\Information Systems}} 
& [5, 18, 21, 30]\\ \\
\hline 
\multirow{5}{*}{\minitab[l]{Advanced Transportation\\Management Systems} }
& [2, 3, 4, 6, 11, 12, 13, 14,\\
& 19, 20, 23, 24, 25, 26, 27,\\
& 28, 29, 31, 32, 33, 34, 35,\\
& 38, 39, 40, 41, 42, 44, 43,\\
& 45, 46 ]\\ 
\hline 
\multirow{1}{*}{\minitab[l]{ITS-Enabled Transportation\\Pricing Systems} }
& [35, 36]\\ \\
\hline 
\multirow{1}{*}{\minitab[l]{Advanced Public\\Transportation Systems} }
& [37]\\ \\ 
\hline 
\multirow{1}{*}{\minitab[l]{Fully Integrated Intelligent\\Transportation Systems} }
& [22, 36]\\ \\ 
\bottomrule
\end{tabular}}
\vskip 3pt
\end{center}

It is observed that most of the current studies fit the category of Advanced Transportation Management Systems, one of the most ``management oriented" areas in ITS. This is not surprising because OCT, as we mentioned previously, is a sharp tool in addressing management issues in engineering. The current adoption situation of OCT simply restates this view in ITS area. This conveys an encouraging message, namely though many management issues in ITS are novel and complex, as a rigorous research method, OCT can still effectively address many of the challenges in this emerging area.

It{\textquoteright}s very interesting to see there is little research in the Advanced Public Transportation Systems area and Fully Integrated Intelligent Transportation Systems. The reason why there are not many studies in these areas so far worth a further investigation in a larger body of journals and conferences. Our conjecture is that there might be less identified management oriented issues in these areas and thus it{\textquoteright}s harder for OCT to play its role. However, this contention needs to be further examined. We will further discuss this in the next section. On the other hand, the lack of studies in these areas may suggest research gaps where OCT could step in and solve a rich body of problems if more management issues are to be found. Next section will summarize the findings we generated and discuss the research gaps and possible future research directions in greater detail.

\section{Discussion, research gaps and possible future research directions}
\subsection{Summary of the findings}
First of all, we found OCT is widely used in ITS research and many studies have been published in mainstream transportation journals especially in Transportation Research Part B: Methodological and Transportation Research Part C: Emerging Technologies. Publications in these top journals showed the effectiveness and validity of OCT in achieving ITS research objectives.

Secondly, we found ITS problems on land road are the main target of OCT. This is consistent with the current state-of-the-art of ITS as we mentioned in section 2.3. We also found reducing traffic congestion is the main focus of current optimal control based literature. This phenomenon implies the usage of optimal control in the extant literature is focusing on one of the main objectives of ITS [1].

Thirdly, we found current studies in ITS which adopt OCT mainly addresses problems within the scope of Advanced Transportation Management Systems. While this confirms the effectiveness of OCT in solving management oriented issues in ITS, the lack of studies in other ITS areas may signify the research gaps and bring research opportunities. We shall discuss this in the next section.

Now we can answer our first research question. In the current literature, OCT contributes to ITS research mainly by addressing various management oriented research problems within with reducing traffic congestion, improving operation and maintenance efficiency and effectiveness are the major focus.

\subsection{Research gaps and possible future research directions}
As we previously discussed, the lack of current studies in some ITS research areas other than Advanced Transportation Management Systems may be due to the less identified management issues in these areas. We argue that many potentially important management oriented issues could be discovered in these areas. For instance, in the area of Advanced Public Transportation Systems, many topics such as fleet management of buses, energy management of public transportations, etc. has considerable amount of unsolved research problems. In the area of Fully Integrated Intelligent Transportation Systems, even more potential management related issues would be raised especially in the upcoming era of Internet of Things (IoT), vehicle telemetry and ubiquitous computing. It could be expected that in the near future vehicle to vehicle (V2V) communication and vehicle to infrastructure (V2I) communication will be prevalent [1, 47, 48, 49]. There will be enormous management and control related research questions such as ``How to manage V2V enabled fleet to minimize the travel time?" and ``How to take advantage of interconnected vehicles and traffic signs to reduce traffic congestion?" and so on. Since OCT already rigorously contributed to ITS by solving many important and novel research problems, it{\textquoteright}s reasonable to propose OCT as an important candidate in solving new management and control related issues ITS.

Therefore, we believe the current research gaps in the areas such as Fully Integrated Intelligent Transportation Systems and Advanced Public Transportation Systems would create research opportunities for OCT researchers if they dive deeper in this exciting and increasingly more interconnected transportation world.

\section{Concluding remarks}
We selectively reviewed 41 high quality OCT based research articles published in 11 top transportation journals. We aimed to answer two research questions and identify the research gaps and possible future research directions of using OCT in ITS. We categorized the research articles by four different ways and found OCT is widely used in current ITS research and it is able to address important topics in ITS. Characteristics of these studies were revealed and discussed. We also examined the literature and identified the current research gaps and pointed out the possible future research directions in this area.

\vspace{.5\baselineskip} {\selectfont\scriptsize                        % Please supply authors' biographies here.
%\lettrine[image=true, lines=9, lhang=0, loversize=0, %
%          lraise=0, findent=2mm, nindent=0mm]
%{Author}
{\bf Jimmy SJ. REN} is a Ph.D. candidate of Department of Information Systems, City University of Hong Kong. He received his BEng degree in Software Engineering from East China Normal University. His research interests include optimal control theory, machine learning and Intelligent Transportation Systems. E-mail: justhavealittlefaith@gmail.com.
\par}

\vspace{.5\baselineskip} {\selectfont\scriptsize
%\lettrine[image=true, lines=9, lhang=0, loversize=0, %
%          lraise=0, findent=2mm, nindent=0mm]
%{Author}
{\bf Wei WANG} received the Ph.D. degree in Information Systems from City University of Hong Kong in 2012 and he received the B.S. degree in Electrical Engineering and the M.S. degree in Microwave Engineering from Peking University, Beijing, China in 2005 and 2008 respectively.  
He is currently a research assistant in Department of Information Systems of City University of Hong Kong. His current research interests include machine learning and Intelligent Transportation Systems. E-mail: wewang8@cityu.edu.hk.
\par}

\vspace{.5\baselineskip} {\selectfont\scriptsize
%\lettrine[image=true, lines=9, lhang=0, loversize=0, %
%          lraise=0, findent=2mm, nindent=0mm]
%{Author}
{\bf Stephen Shaoyi LIAO} is Professor and director of Advanced Transportation Information Systems (ATIS) Research Center of the City University of Hong Kong. He got a bachelor degree from Beijing University and earned a Ph.D. from University of Aix-Marseille III and Institute of France Telecom.  His research articles have been published in various academic journals like MIS Quarterly, IEEE transactions, Communications of the ACM, Transportation Research, Decision Support Systems, Information Science, Computer Software and Applications. He actively works on the areas of Intelligent Transportation Systems and currently the project coordinator of several Innovation Technology Funds projects on ITS funded by Hong Kong SAR government.  His current research interests include use of Information Technology in mobile commerce applications and intelligent business systems, especially Intelligent Transportation Systems. E-mail: issliao@cityu.edu.hk.
\par}

\end{document}